\documentclass[epj]{svjour}
\usepackage{graphics}
\usepackage{epsfig}

\newcommand{\Lcal}{{\mathcal L}}
\newcommand{\psib}{\bar{\psi}}

\newcommand{\btau}{\mbox{\mbox{\boldmath$\tau$}}}

\newcommand{\bPhi}{\mbox{\mbox{\boldmath$\Phi$}}}

\newcommand{\bfAm}{\mbox{\mbox{\boldmath$A$}}}

\newcommand{\nuk}{{\nu_{\rm K}}}
\newcommand{\msun}{M_{\odot}}
\newcommand{\pkgr}{{P_{\,\rm K}}}
\newcommand{\okgr}{{\Omega_{\,\rm K}}}
\newcommand{\mevt}{{\rm MeV/fm}^3}
\newcommand{\const}{{\rm const}}
\newcommand{\kFchi}{k_{F_\chi}}
\newcommand{\kFn}{{k_{F_n}}}
\newcommand{\fmmo}{{\rm fm}^{-1}}
\newcommand{\glen}{{\rm G}^{\rm {B180}}_{300}}

\begin{document}

\title{Rotating Neutron Stars}


\author{Fridolin Weber %\inst{1}
\thanks{The research of F.\ Weber is supported by the National Science
Foundation under Grant PHY-0457329, and by the Research Corporation.}
\and Philip Rosenfield%\inst{1}
}                 % Do not remove

\institute{Department of Physics, San Diego State University, 5500
Campanile Drive, San Diego, CA 92182-1233, USA}

\date{Received: date / Revised version: date}

\abstract{Because of the tremendous densities that exist in the cores of
  neutron stars, a significant fraction of the matter in the cores of such
  stars is likely to exist in the form of hyperons. Depending on spin
  frequency, the hyperon content changes dramatically in rotating neutron
  stars, as discussed in this paper.
 % end of PACS codes
} %end of abstract

\maketitle

\section{Introduction}
\label{intro}

\begin{figure}[tb] 
\begin{center}
\includegraphics*[width=0.40\textwidth,angle=-90]{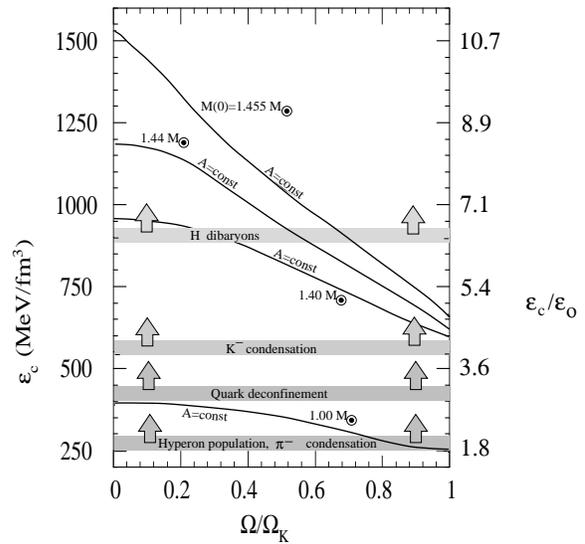}
\caption[]{Change of central density with rotational neutron star frequency
  \cite{weber99:book}.  $\epsilon_0 = 140~\mevt$ denotes the density of nuclear
  matter, $\okgr$ is the Kepler frequency, and $M(0)$ is the star's mass at
  zero rotation.}
\label{fig:ec1445fig}
\end{center}
\end{figure}

Rotating neutron stars are called pulsars. Three distinct classes of pulsars
are currently known. These are (1) rota-tionpowered pulsars, where the loss of
rotational energy of the star powers the emitted electromagnetic radiation,
(2) accretion-powered (X-ray) pulsars, where the gravitational potential
energy of the matter accreted from a low-mass companion is the energy source,
and (3) magnetars, where the decay of a ultra-strong magnetic field powers the
radiation. The matter in the cores of rotating neutron stars is compressed to
ultra-high densities that may be more than an order of magnitude greater than
the density of atomic nuclei.  This makes (rotating) neutron stars superb
astrophysical laboratories for a wide range of fascinating physical studies
\cite{glen97:book,weber99:book,blaschke01:trento}. This includes the physics
of hyperons in cold ultra-dense matter, whose thresholds, according to model
calculations, are easily reached in the cores of neutron stars (see Fig.\
\ref{fig:ec1445fig}), depending on the mass and rotational frequency of a
neutron star.  The most rapidly rotating, currently known neutron star is
pulsar PSR J1748-2446ad, which rotates at a period of 1.39~ms (which
corresponds to a rotational frequency of 719~Hz) \cite{hessels06:a}. It is
followed by PSRs B1937+21 \cite{backer82:a} and B1957+20 \cite{fruchter88:a}
whose rotational periods are 1.58 ms (633~Hz) and 1.61~ms (621~Hz),
respectively. Finally, we mention the recent discovery of X-ray burst
oscillations from the neutron star X-ray transient XTE J1739--285
\cite{kaaret06:a}, which could suggest that XTE J1739--285 contains the most
rapidly rotating neutron star yet discovered.
Rapid rotation changes the structure and composition of neutron stars
dramatically, and leads to novel phenomena like frame dragging (Lense Thirring
effect), as discussed in this paper. 

\section{Stellar structure equations}

Since neutron stars are objects of highly compressed matter, the geometry of
spacetime is changed dramatically from flat space by these objects. Neutron
star models are thus to be computed from Einstein's field equations of
general relativity ($\mu, \nu$=0,1,2,3),
\begin{eqnarray}
  G_{\mu\nu} \equiv R_{\mu\nu} - {{1}\over{2}} g_{\mu\nu} R = 8 \pi
  T_{\mu\nu}(\epsilon,P(\epsilon)) \, ,
\label{eq:intro.1}
\end{eqnarray} which couples Einstein's curvature tensor, $G_{\mu\nu}$, to the
energy--momentum density tensor, $T_{\mu\nu}$, of the stellar
matter. The quantities $R_{\mu\nu} \equiv \Gamma_{\mu\sigma, \,
  \nu}^\sigma - \Gamma_{\mu\nu, \, \sigma }^\sigma +
\Gamma_{\kappa\nu}^\sigma \, \Gamma_{\mu\sigma}^\kappa -
\Gamma_{\kappa\sigma}^\sigma \, \Gamma_{\mu\nu}^\kappa$, $g_{\mu\nu}$,
and $R \equiv R_{\mu \nu} g^{\mu \nu}$ in denote the Ricci tensor,
metric tensor and scalar curvature, respectively
\cite{weber99:book}. (Commas followed by a Greek letter denote
derivatives with respect to space-time coordinates, e.g. ${,_\nu} =
{{\partial}/{\partial x^\nu}}$.)  The Christoffel symbols
are defined as $\Gamma_{\mu\nu}^{\sigma} = (1/2) g^{\sigma\lambda}$
$( g_{\mu\lambda,\, \nu} + g_{\nu\lambda,\, \mu } - g_{\mu\nu, \,
\lambda})$.  Theories of superdense matter enter in Eq.\
(\ref{eq:intro.1}) through the energy--momentum tensor,
\begin{eqnarray}
  T_{\mu\nu} = u_\mu \, u_\nu \, (\epsilon + P) +
   g_{\mu\nu} P \, ,
\label{eq:85.7}
\end{eqnarray}
which contains the equation of state (EoS), i.e.\ pressure as a function on
energy density, $P(\epsilon)$, of the stellar matter.  The quantities $u_\mu$
and $u_\nu$ in Eq.\ (\ref{eq:85.7}) are four-velocities, defined as $u_\mu = d
x_\mu / d\tau$ and $u_\nu = d x_\nu / d\tau$, where $d\tau^2 = - d s^2$. These
velocities are the components of the macroscopic velocity of the stellar
matter with respect to the actual coordinate system that is being used.

\subsection{Non-rotating stars}

Non-rotating neutron stars are spherically symmetric. The metric of such
stars thus dependes only on the radial coordinate and is given by
\begin{eqnarray}
  d s^2 = - e^{2\,\Phi} \, d t^2 + e^{2\,\Lambda} \, d r^2 +
  r^2 \, d\theta^2 + r^2 \, {\rm sin}^2\theta \, d\phi^2\, ,
\label{eq:15.20}
\end{eqnarray} 
where $\Phi$ and $\Lambda$ are the radially varying metric functions.
From Eq.\ (\ref{eq:15.20}) it follows that the components of the
metric tensor are given by
\begin{equation}
%    g(dn,dn)   [1, 1]
g_{t t} = - \, e^{2\,\Phi} \, , ~
%    g(dn,dn)   [2, 2]
g_{r r} = e^{2\,\Lambda} \, , ~
%    g(dn,dn)   [3, 3]
g_{\theta \theta} =  {r}^{2} \, , ~
%    g(dn,dn)   [4, 4]
g_{\phi \phi} =  {r}^{2} \sin^2\!\theta \, ,
\label{eq:15.34}
\end{equation}
so that the only non-vanishing Christoffel symbols are
\begin{eqnarray}
%    Chr(dn,dn,up)   [1, 1, 2]  
&&\Gamma_{t t}^r = 
 {e^{2\Phi-2\Lambda}} \Phi' \, , 
%    Chr(dn,dn,up)   [1, 2, 1]
\Gamma_{t r}^t =  \Phi' \, , 
%    Chr(dn,dn,up)   [2, 2, 2]
\Gamma_{r r}^r = \Lambda'  \, ,
%    Chr(dn,dn,up)   [2, 3, 3]
\Gamma_{r \theta}^\theta = {r}^{-1}  \, ,  
\nonumber \\ 
%    Chr(dn,dn,up)   [2, 4, 4]
&&\Gamma_{r \phi}^\phi =  {r}^{-1}  \, , 
%    Chr(dn,dn,up)   [3, 3, 2] 
\Gamma_{\theta \theta}^r = - r e^{-2\Lambda}  \, , 
%    Chr(dn,dn,up)   [3, 4, 4]
\Gamma_{\theta \phi}^\phi =  {\frac {\cos\theta}{\sin\theta}} \, , 
\nonumber \\
%    Chr(dn,dn,up)   [4, 4, 2]
&&\Gamma_{\phi \phi}^r = - r \sin^2 \!\theta \,
e^{-2 \Lambda}  \, , 
%    Chr(dn,dn,up)   [4, 4, 3]
\Gamma_{\phi \phi}^\theta = -\sin\theta \, \cos\theta \, .
\label{eq:15.54c}
\end{eqnarray} 
Subsituting Eqs.\ (\ref{eq:15.34}) and (\ref{eq:15.54c}) into
(\ref{eq:intro.1}) and using $T_{{\mu\nu}; \mu} = 0$ (the semicolon
denotes covariant differentiation) leads to the
Tolman--Oppenheimer--Volkoff (TOV) equation \cite{weber99:book},
\begin{eqnarray}
  {{dP}\over{dr}} = - \, \frac{\epsilon \, m} {r^2} \frac{ \left( 1
+ P / \epsilon \right) \left( 1 + 4 \pi r^3 P/m \right) } {1
- 2 m/r} \; ,
\label{eq:f28}
\end{eqnarray} applicable to compact stellar configurations in 
general relativistic hydrostatic equilibrium.  We use units for which
the gravitational constant and velocity of light are $G=c=1$ so that
the mass of the sun is $\msun = 1.47$ km. The mass contained in a
sphere of radius $r$ is given by $m = 4 \pi \int^r_0 r^2 \epsilon
dr$. Hence the star's total mass follows as $M\equiv m(R)$, where $R$
denotes the star's radius.  The Newtonian limit of Eq.\ (\ref{eq:f28})
\begin{figure}[tb]
\begin{center}
\includegraphics*[width=0.35\textheight]{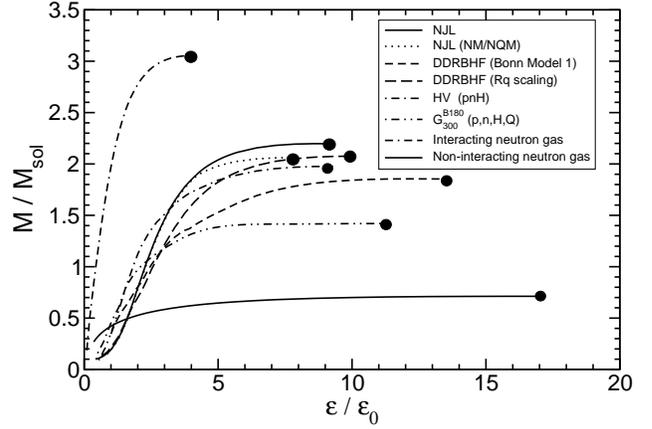}
\caption{Neutron star mass versus central energy density for different
  equations of state of neutron star matter \cite{weber07:a}.}
\label{fig:ec} 
\end{center}
\end{figure} 
is obtained for $P/\epsilon \ll 1$, $4 \pi r^3 P/m \ll 1$, and $2m/r \ll 1$.
General relativity, thus, increases the pressure gradient inside the star
which leads to smaller, more compact stars.  The masses of neutron stars lie
between about one and two solar masses, and their radii are around 10~km.
Thus, $2 M / R \sim 30 - 60 \%$ for neutron stars. Solutions of the TOV
equations are shown in Fig.\ \ref{fig:ec}.

\subsection{Rotating stars}

The structure equations of rotating neutron stars are considerably more
complicated than those of non-rotating neutron stars \cite{weber99:book}.
These complications have their cause in the rotational deformation, that is, a
flattening at the pole accompanied by a radial expansion in the equatorial
direction, which leads to a dependence of the star's metric on the polar
coordinate, $\theta$. Secondly, rotation stabilizes a neutron star against
gravitational collapse. A rotating neutron star can therefore carry more mass
than a non-rotating star.  Being more massive,
\begin{figure}[htb]
\begin{center}
\includegraphics*[width=0.4\textwidth,angle=0,clip]{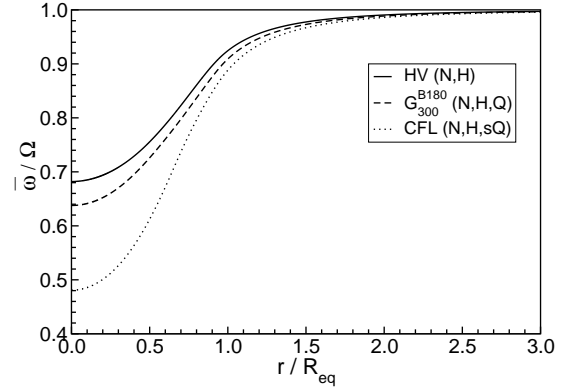}
\caption{Dragging of local inertial frames (Lense-Thirring effect)
  caused by $\sim 1.4 \, \msun$ neutron stars rotating at 2~ms
  \cite{weber05:a}.  The frequency $\bar\omega$ is defined in Eq.\
  (\protect \ref{eq:bar.omega}).}
\label{fig:fd}
\end{center}
\end{figure}
however, means that the geometry of space-time is changed too. This renders
the metric functions of a rotating neutron star frequency dependent.  Finally,
the general relativistic effect of the dragging of local inertial frames
implies the occurrence of an additional non-diagonal term, $g^{t\phi}$, in the
metric tensor $g^{\mu\nu}$. This term imposes a self-consistency condition on
the stellar structure equations, since the degree at which the local inertial
frames are dragged along by the star is determined by the initially unknown
stellar properties like mass and rotational frequency. The covariant
components of the metric tensor of a rotating compact star are thus given by
\cite{weber99:book,friedman86:a}
\begin{eqnarray}
  g_{t t} &=& - e^{2\nu} + e^{2\psi} \omega^2  \, , ~ g_{t \phi}
  = - e^{2\psi} \omega  \, , ~ g_{r r} = e^{2\lambda} \, , ~ \nonumber \\
  g_{\theta \theta} &=& e^{2\mu} \, , ~  g_{\phi \phi} =
  e^{2\,\psi} \, ,
\label{eq:11.4bk} 
\end{eqnarray}
which corresponds to a line element, $ds^2 = g_{\mu\nu} dx^\mu dx^\nu$, of the
form
\begin{eqnarray}
  d s^2 = - e^{2\nu} dt^2 + e^{2\psi} (d\phi - \omega dt)^2 + e^{2\mu}
  d\theta^2 + e^{2\lambda} dr^2 \, .
\label{eq:f220.exact} 
\end{eqnarray} Here each metric function, i.e.\ $\nu$, $\psi$,  $\mu$ and
$\lambda$, as well as the angular velocities of the local inertial
frames, $\omega$, depend on the radial coordinate $r$ and on the
polar angle $\theta$ and, implicitly, on the star's angular velocity
$\Omega$.  Of particular interest is the relative angular frame
dragging frequency, $\bar\omega$, defined as
\begin{equation}
  \bar\omega(r,\theta,\Omega) \equiv \Omega - \omega(r,\theta,\Omega)
  \, ,
\label{eq:bar.omega}
\end{equation} 
which is the angular velocity of the star, $\Omega$, relative to the angular
velocity of a local inertial frame, $\omega$. It is this frequency that is of
relevance when discussing the rotational flow of the fluid inside the star,
since the magnitude of the centrifugal force acting on a fluid element is
governed--in general relativity as well as in Newtonian gravitational
theory--by the rate of rotation of the fluid element relative to a local
inertial frame \cite{hartle67:a}. In contrast to Newtonian theory, however,
the inertial frames inside (and outside) a general relativistic fluid are not
at rest with respect to the distant stars. Rather, the local inertial frames
are dragged along by the rotating fluid. Depending on the internal stellar
constitution, this effect can be quite strong, as shown in Fig.\ \ref{fig:fd}
\begin{figure}[tb]
\begin{center}
\includegraphics*[width=0.40\textwidth]{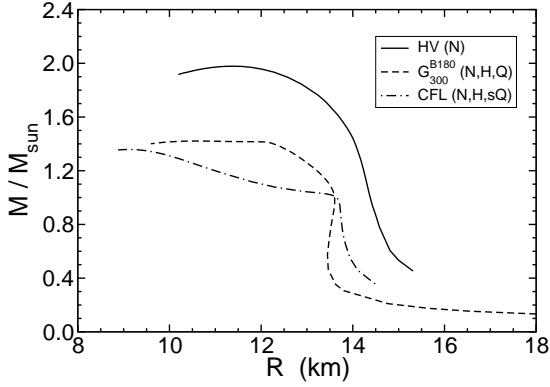} 
\caption[]{Mass--radius relations of non-rotating  neutron stars
          for different EoSs.}
\label{fig:mvsr_0}
\end{center}
\end{figure}
for neutron stars rotating at 2~ms \cite{weber05:a}.  The stellar models are
computed for three different equations of state: (1) HV, which describes
neutron stars made of nucleons (N) and hyperons (H); $\glen$, which describes
neutron star matter in terms of nucleons, hyperons, and quarks (Q); and CFL,
which assumes that the quarks are color superconducting (CFL phase)
\cite{alford03:a}.  For a very compact neutron star, as obtained for the CFL
case, one sees that the local inertial frames at the star's center rotate at
about half the star's rotational frequency, $\omega(r=0) \simeq \Omega/2$.
This value drops to about 15\% for the local inertial frames located at the
star's equator.

\begin{figure}[tb]
\begin{center}
\includegraphics*[width=0.40\textwidth,angle=0,clip]{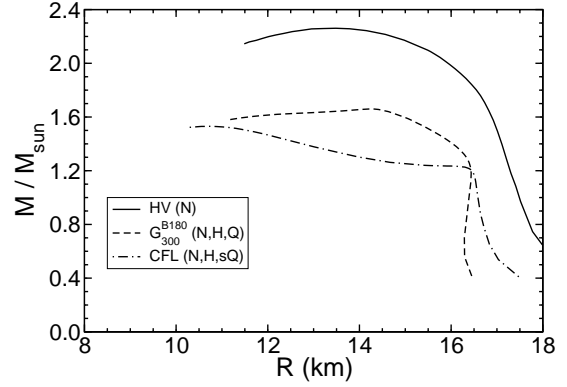}
\caption{Same as Fig.\ \ref{fig:mvsr_0}, but for neutron stars
  rotating at the mass shedding (Kepler) frequency.}
\label{fig:mvsr_K}
\end{center}
\end{figure}

Figures \ref{fig:mvsr_0} and \ref{fig:mvsr_K} show the influence of
rotation on the mass--radius relationship of neutron stars.
For ultrafast rotation at the Kepler frequency, a mass increase up to $\sim
20$\% is obtained, depending on the equation of state. The equatorial radius
increases by several kilometers, while the polar radius get smaller by several
kilometers. The ratio between both radii is around 2/3, except for rotation
close to the Kepler frequency.

\subsection{Limiting rotational periods}\label{sssec:grav}

No simple stability criteria are known for rapidly rotating stellar
configurations in general relativity. However, an absolute limit on rapid
rotation is set by the onset of mass shedding from the equator of a rotating
star. The corresponding rotational frequency is known as the Kepler frequency,
$\okgr$. In classical mechanics, the expression for the Kepler frequency,
determined by the equality between the centrifugal force and gravity, is
readily obtained as $\okgr = \sqrt{M/R^3}$. In order to derive the general
relativistic counterpart of this relation, one applies the extremal principle
to the circular orbit of a point mass rotating at the star's equator.  Since
$r=\theta=\const$ for a point mass there, one has $d r=d\theta=0$.  The line
element (\ref{eq:f220.exact}) then reduces to $d s^2 = ( e^{2\nu} - e^{2\psi}
(\Omega - \omega)^2 ) \, d t^2$.  Substituting this expression into $J \equiv
\int^{s_2}_{s_1} d s$, where $s_1$ and $s_2$ refer to points located at
\begin{figure}[tb]
\begin{center}
\includegraphics*[width=0.4\textwidth,angle=0,clip]{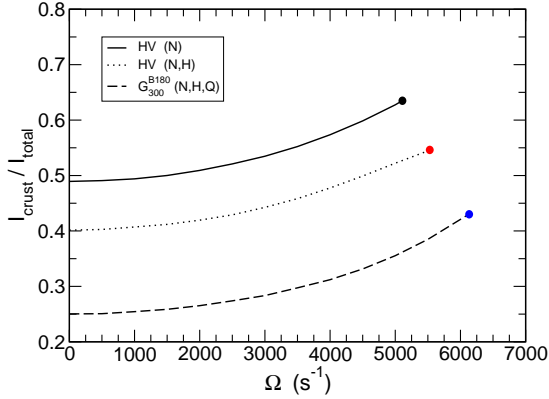}
\caption{Moment of inertia of several sample neutron stars.}
\label{fig:moi}
\end{center}
\end{figure}
that particular orbit for which $J$ becomes extremal, gives
\begin{eqnarray}
  J = \int_{s_1}^{s_2}\! d t \, \sqrt{e^{2\,\nu} - e^{2\,\psi}
    \, (\Omega - \omega)^2} \, .
\label{eq:12.2bk}
\end{eqnarray} Applying  the extremal condition $\delta J=0$ to
Eq.\ (\ref{eq:12.2bk}) and noticing that $V = e^{\psi-\nu} \, (\Omega
- \omega)$ then leads to
\begin{eqnarray}
  \psi_{,r} \, e^{2\,\nu} \, V^2 - \omega_{,r} \, e^{\nu+\psi}\, V -
  \nu_{,r} \, e^{2\nu} = 0 \, .
\label{eq:12.6bk}
\end{eqnarray} 
This relation constitutes a quadratic equation for the orbital velocity $V$ of
a particle at the star's equator. One thus obtains for the Kepler frequency
$\okgr$ the final relation \cite{weber99:book}
\begin{eqnarray}
  \okgr = \omega +\frac{\omega_{,r}} {2\psi_{,r}} + e^{\nu -\psi} \sqrt{
    \frac{\nu_{,r}} {\psi_{,r}} + \Bigl(\frac{\omega_{,r}}{2
      \psi_{,r}} e^{\psi-\nu}\Bigr)^2 } \, , 
\label{eq:okgr}  
\end{eqnarray} 
which is to be solved self-consistently at the equator of a rotating neutron
star. The general relativistic Kepler period follows from Eq.\ (\ref{eq:okgr})
as $\pkgr = {{2 \pi} / {\okgr}}$. For typical neutron star matter equations of
state, the Kepler period obtained for $1.4\, \msun$ neutron stars scatters
around 1~ms. One exception to this are strange quark matter stars. These are
self-bound and, thus, tend to possess smaller radii than conventional neutron
stars, which are bound by gravity only. Because of their smaller radii,
strange stars can withstand mass shedding down to periods of around 0.5~ms
\cite{glen92:crust,glen92:limit}. The CFL neutron star discussed just above is
a quark-hybrid star and as such bound by gravity only. The dense quark core in
the center of this neutron star implies a relatively large binding energy of
$0.12 \, \msun$, leading to a rather low mass shedding period of 0.7~ms.
In closing this section, we introduce the moment of inertia of a rotating
neutron star described by the metric in Eq.\ (\ref{eq:f220.exact}). For such
stars the moment of inertia is given by
\begin{eqnarray}
  I = 2 \pi \int_0^\pi \! d\theta \int_0^{R(\theta)} d r \,
  e^{\lambda+\mu+\nu+\psi}  {{\epsilon + P}\over{e^{2\nu -
        2\psi} - \bar\omega^2}} \, {{\bar\omega}\over{\Omega}}
  \, . 
\label{eq:11.71bk} 
\end{eqnarray}
Figure~\ref{fig:moi} shows that the crustal fraction of the moment of
inertia of a neutron star may be around 50\% smaller
if the star contains a very soft phase of matter like CFL quark
matter. This may be of relevance for pulsar glitch models and the
modeling of the post-glitch behavior of pulsars.

\section{Composition of neutron star matter}

A vast number of models for the equation of state of neutron star
matter has been derived in the literature over the years. These models
can roughly be classified as follows:
\begin{itemize}
\item Thomas-Fermi based models \cite{myers95:a,strobel97:a}
\item Schroedinger-based models (e.g.\
variational approach, Monte Carlo techniques, hole line expansion
(Brueckner theory), coupled cluster method, Green function method)
\cite{heiselberg00:a,pandharipande79:a,wiringa88:a,akmal98:a}
\item Relativistic field-theoretical treatments (relativistic mean field
(RMF), Hartree-Fock (RHF), standard Brueckner-Hartree-Fock (RBHF),
density dependent RBHF (DD-RBHF)
\cite{glen97:book,lenske95:a,fuchs95:a,typel99:a,hofmann01:a,niksic02:a,ban04:a}
\item Nambu-Jona-Lasinio (NJL) models 
\cite{buballa05:a,blaschke05:a,rischke05:a,abuki06:a,lawley06:a,lawley06:b}
\item Chiral SU(3) quark mean field model \cite{wang05:a}.
\end{itemize}
Neutron star masses computed for some of these models are shown in Fig.\
\ref{fig:ec}.

\subsection{Relativistic nuclear field-theoretical models}

Relativistic nuclear field-theoretical models
\cite{glen97:book,weber99:book,lenske95:a,fuchs95:a,typel99:a,hofmann01:a,%
  niksic02:a,ban04:a} are based on  Lagrangians of the form $\Lcal = \Lcal_{B}
+ \Lcal_{M} + \Lcal_{int} + \Lcal_{lept}$, where
\begin{eqnarray}
  \Lcal_{B} &=& \sum_B \psib_B \left( i\gamma_\mu\partial^\mu - m_B \right) \psi_B \, ,
\label{eq:LB} \\ 
\Lcal_{M} &=&\frac{1}{2} \sum_{i=\sigma,\delta}
\left(\partial_\mu\Phi_i\partial^\mu\Phi_i - m_i^2\Phi_i^2\right) 
\nonumber \\  &&-
\frac{1}{2} \sum_{\kappa=\omega,\rho} \Bigl( \frac{1}{2}
F^{(\kappa)}_{\mu\nu}  F^{(\kappa)\mu\nu}  
 - m_\kappa^2 A^{(\kappa)}_\mu
A^{(\kappa)\mu} \Bigr) \, , \label{eq:Lagrangian} \\
\Lcal_{int} &=&\psib\hat{\Gamma}_{\sigma}(\psib,\psi)\psi\Phi_{\sigma}
- \psib\hat{\Gamma}_{\omega}(\psib,\psi)\gamma_{\mu}\psi A^{(\omega)
  \mu} \nonumber \\ && + \psib\hat{\Gamma}_{\delta}(\psib,\psi) 
{\btau} \psi {\bPhi}_{\delta} - \psib\hat{\Gamma}_{\rho}
(\psib,\psi)\gamma_{\mu} {\btau} \psi {\bfAm}^{(\rho)\mu} \,
. \label{eq:Lint}
\end{eqnarray}
Here, $\Lcal_B$ and $\Lcal_M$ are the free baryonic and the free mesonic
Lagrangians, respectively, and interactions are described by $\Lcal_{int}$,
where $ F^{(\kappa)}_{\mu\nu} = \partial_\mu A_\nu^{(\kappa)} - \partial_\nu
A_\mu^{(\kappa)}$ is the field strength tensor of one of the vector mesons
($\kappa= \omega, \rho$).  In the case of RMF, RHF and RBHF the meson-baryon
vertices $\hat\Gamma_\alpha$ ($\alpha=\sigma,\omega,\delta, \rho$) are
density-independent quantities which are given by expressions like
$\hat\Gamma_\sigma = i g_\sigma$ for the scalar $\sigma$ meson,
$\hat\Gamma^\mu_\omega = g_\omega \gamma^\mu + (i/2) (f_\omega / 2 m)
\partial_\lambda [\gamma^\lambda,\gamma^\mu]$ for $\omega$ mesons, etc.
\cite{weber99:book}. In the framework of the DD-RBHF scheme, the meson-baryon
vertices $\hat\Gamma_\alpha$ depend on the baryon field operators $\psi$
\cite{hofmann01:a}. The field equations that follow from Eqs.\
(\ref{eq:LB})--(\ref{eq:Lint}) have the mathematical form
\begin{eqnarray}
  ( i \gamma^\mu\partial_\mu - m_B ) \psi_B(x) &=&
  \sum_{M= i, \kappa}  \, M(x) \, \hat\Gamma_M \, 
  \psi_B(x) \, , \\ ( \partial^\mu\partial_\mu + m^2_\sigma)
  \sigma(x) &=& \sum_B \, \bar\psi_B(x) \,
  \hat\Gamma_\sigma  \, 
  \psi_B(x) \, ,
\end{eqnarray}
plus similar equations for the other mesons \cite{weber99:book,hofmann01:a}.

\subsection{Hyperons and baryon resonances}

\begin{figure}[tb]
\begin{center}
\includegraphics*[width=0.40\textwidth]{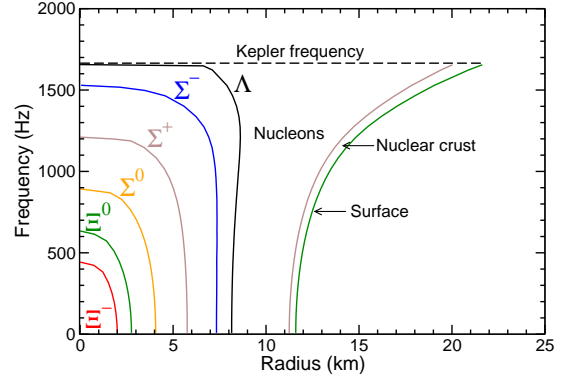}
\caption[]{Hyperon composition of a rotating neutron star in equatorial
  direction. (DD-RBHF calculation performed for Bonn (model 1) potential,
  non-rotating star mass is $1.70\, \msun$.)}
\label{fig:bonn_eq}
\end{center}
\end{figure}

\begin{figure}[tb]
\begin{center}
\includegraphics*[width=0.40\textwidth]{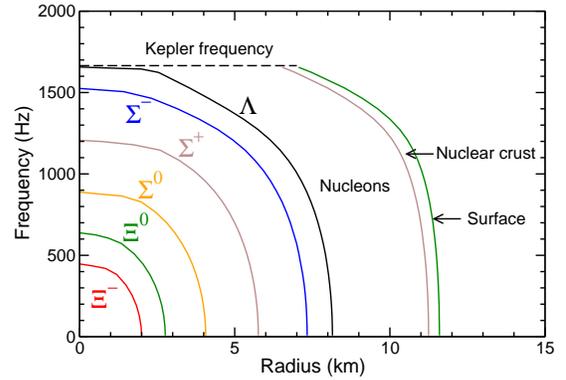} 
\caption[]{Same as Fig.\ \ref{fig:bonn_eq}, but in polar direction.}
\label{fig:bonn_po}
\end{center}
\end{figure}

\begin{figure}[tb]
\begin{center}
\includegraphics*[width=0.40\textwidth]{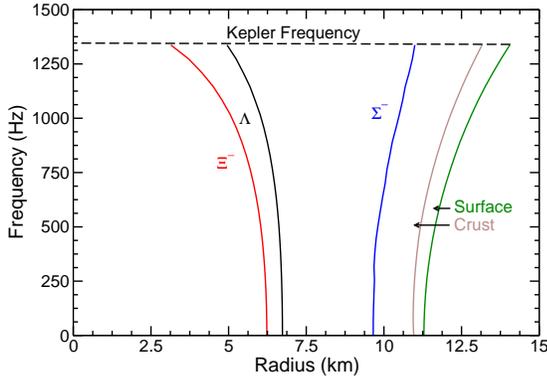}
\caption[]{Hyperon composition of a rotating neutron star in equatorial
  direction. (DD-RBHF calculation performed for Groningen potential,
  non-rotating star mass is $1.60\, \msun$.)}
\label{fig:gron_eq}
\end{center}
\end{figure}

\begin{figure}[tb]
\begin{center}
\includegraphics*[width=0.40\textwidth]{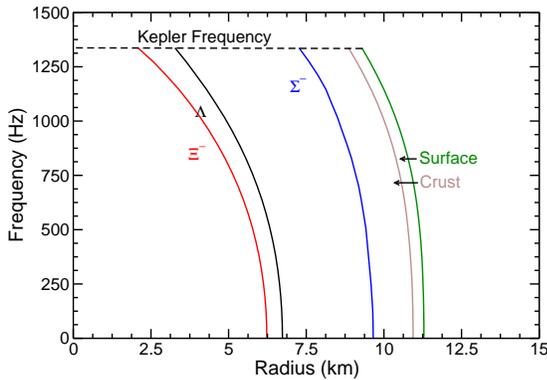} 
\caption[]{Same as Fig.\ \ref{fig:bonn_eq}, but in polar direction.}
\label{fig:gron_po}
\end{center}
\end{figure}

At the densities in the interior of neutron stars, the neutron chemical
potential, $\mu^n$, is likely to exceed the masses, modified by interactions,
of $\Sigma,~ \Lambda$ and possibly $\Xi$ hyperons
\cite{weber99:book,glen85:b}. Hence, in addition to nucleons, neutron star
matter may be expected to contain significant populations of strangeness
carrying hyperons \cite{glen85:b}.  The thresholds of the lightest baryon
resonances ($\Delta^-, \Delta^0, \Delta^+, \Delta^{++}$) are reached for
\begin{figure}[htb]
\begin{center}
\includegraphics*[width=0.40\textwidth]{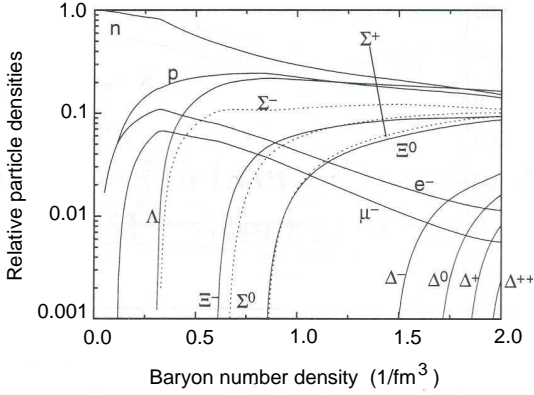}
\caption[]{Composition of neutron star matter in RMF.}
\label{fig:coldnsm}
\end{center}
\end{figure}
relativistic mean-field (RMF) calculations at densities which correspond to
unstable neutron stars. This is different for relativistic
Brueckner-Hartree-Fock (RBHF) calculations where $\Delta$'s appear
\begin{figure}[htb]
\includegraphics*[width=0.35\textwidth]{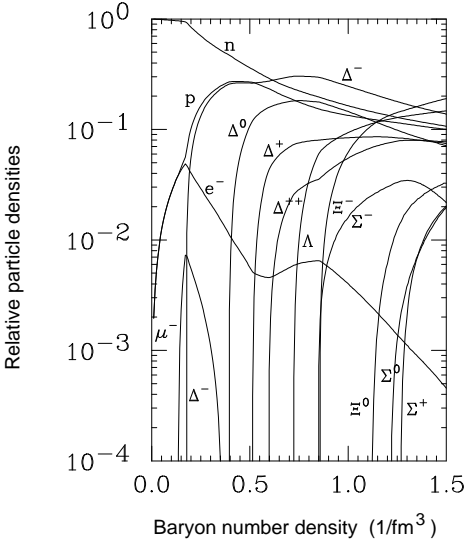}
\caption[]{Same as Fig.\ \ref{fig:coldnsm}, but computed in RBHF
  \protect{\cite{huber98:a}}.}
\label{fig:9.Hub97a}
\end{figure}
rather abundantly in stable neutron stars \cite{huber98:a}, compare Figs.\
\ref{fig:coldnsm} and \ref{fig:9.Hub97a}. Depending on the star mass and
rotational frequency, the total hyperon population in neutron stars can be
very large \cite{glen85:b}, which is illustrated graphically in Figs.\
\ref{fig:bonn_eq}--\ref{fig:gron_po} for rotating neutron stars based on
equations of state computed in the framework of the DD-RBHF formalism. The
stars shown in these figures have rotational frequencies ranging from zero to
the mass shedding frequency, $\nuk$.  Pure neutron matter, therefore,
constitutes an excited state relative to hyperonic matter which would quickly
transform via weak reactions like
\begin{equation}
n \rightarrow p + e^- + {\bar{\nu}}_e 
\label{eq:cnp}
\end{equation} to the lower energy state. The chemical 
potentials associated with reaction (\ref{eq:cnp}) in equilibrium obey the
relation
\begin{equation}
\mu^n = \mu^p + \mu^{e^-} \, ,
\label{eq:mun}
\end{equation} 
where $\mu^{\bar\nu_e}=0$ since the mean free path of (anti) neutrinos is much
smaller than the radius of neutron stars. Hence (anti) neutrinos do not
accumulate inside neutron stars. This is different for hot proto-neutron stars
\cite{prakash97:a}.  Equation~(\ref{eq:mun}) is a special case of the general
relation
\begin{equation}
\mu^\chi = B^\chi \mu^n - q^\chi \mu^{e^-} \, , 
\label{eq:mui}
\end{equation} 
which holds in any system characterized by two conserved charges. These are in
the case of neutron star matter electric charge, $q^\chi$, and baryon number
charge, $B^\chi$. Application of Eq.\ (\ref{eq:mui}) to the $\Lambda$ hyperon
($B^\Lambda=1$, $q^\Lambda=0$), for instance, leads to $ \mu^\Lambda = \mu^n$.
Ignoring particle interactions, the chemical potential of a relativistic
particle of type $\chi$ is given by $\mu^\chi = \omega(\kFchi) \equiv
\sqrt{m_\chi^2 + \kFchi^2}$, where $\omega(\kFchi)$ is the single-particle
energy of the particle and $\kFchi$ its Fermi momentum. One thus obtains
\begin{equation}
  \kFn \geq \sqrt{m_\Lambda^2 - m_n^2} \simeq 3~\fmmo \Rightarrow
    n \equiv { {\kFn^3}\over{3 \pi^2} } \simeq 2 n_0 \, ,
\label{eq:kFn}
\end{equation} 
where $m_\Lambda=1116$~MeV and $m_n=939$~MeV was used. That is, if
interactions among the particles are ignored, neutrons are replaced
with $\Lambda$'s in neutron star matter at densities as low as two
times the density of nuclear matter. This result is only slightly
altered by the inclusion of particle interactions \cite{glen85:b}.
Densities of just $\sim 2 n_0$ are easily reached in the cores of
neutron stars. Neutron stars may thus be expected to contain considerable
populations of $\Lambda$'s, $\Sigma$'s and $\Xi$'s, as confirmed by the
outcome of DD-RBHF calculations shown graphically above.  Depending on the
star's mass, the total hyperon population can be very large \cite{glen85:b}.

%
% BibTeX users please use
%\bibliography{rlist}
%\bibliographystyle{unsrt}

\end{document}